# The electronic phase diagram of the $LaO_{1-x}F_xFeAs$ superconductor

H. Luetkens[1], H.-H. Klauss[2*], M. Kraken[3], F. J. Litterst[3], T. Dellmann[2,3], R. Klingeler[4], C. Hess[4], R. Khasanov[1], A. Amato[1], C. Baines[1], M. Kosmala[5], O.J. Schumann[5], M. Braden[5], J. Hamann-Borrero[4], N. Leps[4], A. Kondrat[4], G. Behr[4], J. Werner[4] and B. Büchner[4]

[1]*Laboratory for Muon-Spin Spectroscopy, Paul Scherrer Institut, CH-5232 Villigen PSI, Switzerland.* [2]*Institut für Festkörperphysik, TU Dresden, D-01069 Dresden, Germany.* [3]*Institut für Physik der Kondensierten Materie, TU Braunschweig, D-38106 Braunschweig, Germany.* [4] *Leibniz-Institut für Festkörper- und Werkstoffforschung (IFW) Dresden, D-01171 Dresden, German,* [5] *II. Phyikalisches Institut, Universität zu Köln D-50937, Germany.* [*]*e-mail: h.klauss@physik.tu-dresden.de.*

**The competition of magnetic order and superconductivity is a key element in the physics of all unconventional superconductors, e.g. in high-transition-temperature cuprates [1], heavy fermions [2] and organic superconductors[3]. Here superconductivity is often found close to a quantum critical point where long-range antiferromagnetic order is gradually suppressed as a function of a control parameter, e.g. charge carrier doping or pressure. It is believed that dynamic spin fluctuations associated with this quantum critical behaviour are crucial for the mechanism of superconductivity. Recently high-temperature superconductivity has been discovered in iron-pnictides providing a new class of unconventional superconductors[4,5,6]. Similar to other unconventional superconductors the parent compounds of the pnictides exhibit a magnetic ground state[7,8] and superconductivity is induced upon charge carrier doping. In this Letter the**

**structural and electronic phase diagram is investigated by means of x-ray scattering, µSR and Mössbauer spectroscopy on the series $LaO_{1-x}F_xFeAs$. We find a discontinuous first-order-like change of the Néel temperature, the superconducting transition temperature and of the respective order parameters. Our results strongly question the relevance of quantum critical behaviour in iron-pnictides and prove a strong coupling of the structural orthorhombic distortion and the magnetic order both disappearing at the phase boundary to the superconducting state.**

The iron-pnictides $RO_{1-x}F_xFeAs$ have a layered crystal structure with alternating FeAs and RO sheets (R=La or Rare Earth), where the Fe atoms are arranged on a simple square lattice [4]. Theoretical studies reveal a two-dimensional electronic structure [13] and conductivity takes place mainly in the FeAs layers while the RO layers provide the charge reservoir. In LaOFeAs a transition from tetragonal to orthorhombic lattice structure at ~160 K [7,15] is present and a long range spin density wave (SDW) antiferromagnetic order with strongly reduced ordered moment has been observed in neutron scattering, Mössbauer spectroscopy and muon spin relaxation (µSR) experiments below 134 K [7,8,16,17]. Superconductivity emerges when doping a magnetic mother compound with electrons or holes and thereby suppressing the magnetic order[7,12]. This suggests an interesting interplay between magnetism and superconductivity and recent theoretical work proposes that magnetic fluctuations associated with a quantum critical point are essential for superconductivity in the electron doped superconductors $RO_{1-x}F_xFeAs$ [14]. Experimentally, the exact nature of the transition from magnetism to superconductivity in iron-pnictides is not well established. For example, in $CeO_{1-x}F_xFeAs$ a gradual suppression of the magnetic order and a subsequent second order like appearance of superconductivity was observed as a function of doping[9]. In contrast, in $SmO_{1-x}F_xFeAs$ both order parameters coexist microscopically in a small doping range of the phase diagram[10], while in $Ba_{1-x}K_xFe_2As_2$ compounds a phase separation into magnetic and superconducting regions has been observed[11].

We report µSR experiments, $^{57}$Fe Mössbauer spectroscopy and XRD measurements on a series of 11 differently doped polycrystalline samples of $LaO_{1-x}F_xFeAs$ with x ranging from 0 to 0.20. Our experiments yield information on both, the doping dependence of the transition temperatures and the respective order parameters. In contrast to $RO_{1-x}F_xFeAs$ with magnetic rare earth ions R, the La system offers the possibility to solely study the magnetic and superconducting properties of the FeAs electronic system without possible interference effects with the rare earth magnetic moment.

In Fig. 1a we show typical zero field (ZF) µSR spectra between 1.6 and 200 K for x=0.0, 0.04, 0.05, 0.10 and 0.20. For all x up to 0.04 we find a spontaneous muon spin precession proving the long range magnetic order and a 100 % magnetically ordered sample volume below $T_N$ (Fig. 1b). The corresponding Néel temperatures gradually decrease from 134 K for x= 0 to 120 K for x=0.04. The analysis of the ZF spectra for x=0 using a model function which takes into account two inequivalent muon sites and which is appropriate for commensurate magnetic order has been described in detail in reference[8]. For x=0.01 and 0.02 the spectra are very similar to those of the x=0 compound and are analyzed using the same model (see supplement). For x=0.04 we find a stronger relaxation of the muon spin precession (see Fig. 1a). The µSR data are better described by a Bessel function for the main signal fraction, which is typical for incommensurate spin density wave order [19]. A change from commensurate to incommensurate order with increasing x can be expected since an increase of electron doping changes the different Fermi surfaces thereby modifying the length of the AFM nesting vector responsible for the magnetic order [20].

All samples with x≥0.05 show a typical paramagnetic signal with a tiny Gaussian relaxation due to nuclear moments down to the lowest measured temperature of 1.6 K. This relaxation increases slightly with x due to an increasing abundance of the strong nuclear moment of the $^{19}$F isotope. For x=0.05, 30% of the signal shows an additional exponential relaxation of electronic origin below 5 K which is at least one order of magnitude smaller than in the lower doped samples. Due to the weak relaxation and the absence of a coherent muon spin precession this is indicative for spurious and diluted static disordered magnetism in a minor sample volume of ~5% (see supplement). For x≥0.075, our experiments clearly prove the absence of any static magnetic correlations.

The results from the μSR experiments are further corroborated by a series of low temperature Mössbauer spectra (Fig. 1d). They show a magnetic hyperfine splitting in addition to a weak quadrupole splitting only for x=0, 0.01, 0.02 and 0.04. Since the magnetic hyperfine field is proportional to the on-site static magnetic moment of the Fe ions and since transferred dipole and hyperfine fields can be neglected, the absolute value of the ordered moment can be determined. Using a conversion factor of 1 μ$_B$/ 20 T [21] we find 0.25(5) μ$_B$ for x=0 [7] and a slightly reduced value of 0.22(5) μ$_B$ for x=0.04. With increasing x the low-temperature Mössbauer spectra show a moderate increase of the width of the magnetic hyperfine-field distribution consistent with the μSR results. For x=0.05, 0,075 and 0.10 a single absorption line typical for paramagnetic iron in this system is found[18]. Only for x=0.05 we find a gradual increase of the linewidth below ~10K from 0.15(2) mm/s to 0.25(2) mm/s indicating the presence of very dilute magnetic correlations also consistent with the μSR results.

Fig.2 shows a comparison of the structural and electronic order parameters for different doping x. Pure LaOFeAs exhibits a structural phase transition[15] from the tetragonal high-temperature phase (space group P4/nmm) into an orthorhombic low-temperature modification (space group Cmma), with $a_{orth.} \approx b_{orth.} \approx \sqrt{2} \cdot a_{tet.}$ This transition is of ferroelastic character associated with a splitting of the orthorhombic lattice constants $a_{orth}$ and $b_{orth}$. Note that in both phases all Fe-As bond distances are identical but that the FeAs$_4$ tetrahedra do not exhibit the ideal shape. The electronic impact of the tetragonal to orthorhombic transition is to lift the orbital degeneracy at the Fe-sites. In Fig. 2a we show the orthorhombic splittings in LaO$_{1-x}$F$_x$FeAs with x=0.0 and x=0.04 which amount to a comparable value at low temperatures. The x-ray powder diffraction profiles at the (110)$_{tet}$ peak were fitted by a single Gaussian and the corresponding line width is also plotted in Fig 2a. This peak transforms into the (020)$_{orth}$/(200)$_{orth}$ pair in the orthorhombic phase yielding considerable broadening in the single-peak fit. The x=0.04 compound clearly exhibits a step-like increase of this peak width at the tetragonal to orthorhombic transition which is fully absent in the x=0.05 compound. From this and similar observations at the (040)$_{orth}$/(400)$_{orth}$ and other reflection pairs we may unambiguously conclude that the x=0.0 and the x=0.04 samples exhibit the onset of the structural transition at $T_S$=158(3)K and $T_S$=151(3)K, respectively, whereas the x=0.05 compound shows only a weak line width increase in the tetragonal phase down to the lowest temperatures. In Fig. 2b we plot the temperature dependence of the local field at the muon site $B_{muon}$ (determined from the muon spin precession frequency $\omega_\mu = \gamma_\mu B_{muon}$, where $\gamma_\mu/2\pi = 135.5$ MHz T$^{-1}$ is the gyromagnetic ratio for the muon) for x=0, 0.01, 0.02 and 0.04 and the mean hyperfine field obtained in the Mössbauer experiments. The curves for x= 0, 0.01 and 0.02 show a small systematic reduction of $T_N$ only, while for x=0.04 we find both, a reduction of $T_N$ and of the low temperature ordered Fe moment size.

To determine the superconducting volume fraction and the magnetic penetration depth in the system LaO$_{1-x}$F$_x$FeAs we performed field cooled transverse field (TF) μSR experiments in an external field of 700 G on specimen with x=0.05, 0.06, 0.075, 0.10, 0.125, 0.15 and 0.20. In such experiments on polycrystalline samples of an anisotropic type-II superconductor bulk superconductivity is revealed by an additional Gaussian relaxation $\sigma_{sc}$ of the muon precession signal below $T_C$. This additional relaxation arises from the inhomogeneous internal field distribution in the vortex phase of the type-II superconductor [22]. It can be clearly seen in the TF spectra shown in Fig. 1c for samples with x=0.05, 0.10 and 0.20. In an anisotropic superconductor the relaxation rate $\sigma_{sc}$ can be converted into $\lambda_{ab}$, the in-plane magnetic penetration depth, via $\lambda_{ab}$ [nm] = 250 ($\sigma_{sc}$ [MHz])$^{-0.5}$ (see [22,23]). $1/\lambda_{ab}^2$ itself is proportional to the superfluid density $n_s$ devided by the effective mass m* of the charge carriers [24]. The temperature dependence of $1/\lambda_{ab}^2$ for all superconducting samples is depicted in Fig. 2b. For x between 0.05 and 0.10 we find a nearly temperature independent behaviour below $T_c$/3 indicative of a low density of states in the superconducting gap. Up to x=0.125 we find an increase of $T_C$ versus $1/\lambda_{ab}^2$ (T → 0) [24] while for x= 0.125, 0.15 and 0.20 $T_C$ is reduced whereas $1/\lambda_{ab}^2$ (T → 0) is almost constant. In addition we find an enhanced concave curvature of $1/\lambda_{ab}^2$ (T) near $T_C$ which is not observed below x=0.10. This behaviour may be associated with multiband effects[25] or the onset of an inhomogeneous doping and a distribution of superconducting ordering temperatures in the sample. For the samples with x=0.15 and 0.20 we indeed find a non-superconducting volume fraction of 15 % at 1.6 K.

The results of our study are summarized in an electronic phase diagram for the transition temperatures and the order parameters of $LaO_{1-x}F_xFeAs$ in Fig. 3. SDW order with high $T_N$ ~120 to 134 K is identified as the ground state throughout the orthorhombic structural phase. Apparently, as soon as the orthorhombic distortion and the SDW magnetism is suppressed we find a ~100% superconducting volume fraction. The superconducting order parameter $1/\lambda_{ab}^2$ (T → 0) increases with x and shows a pronounced maximum around x=0.08 and 0.10. The steplike behaviour of the magnetic as well as the superconducting order parameter between x= 0.04 and 0.05 suggest that the key element for superconductivity in this system is the suppression of the orthorhombic distortion together with the static magnetic order rather than the moderate increase of the charge carrier density by the Fluorine doping. In contrast, in high-$T_C$ cuprates like $La_{2-x}Sr_xCuO_4$ the superconducting order parameter increases linearly from zero as superconductivity appears for increasing x [26]. Note however, that in Nd- and Eu-doped $La_{2-x}Sr_xCuO_4$ a first order transition between stripe magnetism and superconductivity can be induced for large hole doping x ~ 0.12 by a structural lattice distortion from orthorhombic to tetragonal symmetry [27,28]. In contrast to our findings for $LaO_{1-x}F_xFeAs$, a recent neutron scattering study of the doping dependence of the orthorhombic distortion and the magnetic SDW order in $CeO_{1-x}F_xFeAs$ [9] has concluded a gradual suppression of $T_N$ as a function of x within the orthorhombic phase. However, subsequent µSR experiments on the magnetically ordered composition close to the phase boundary (x=0.06) reveals an increased magnetic ordering temperature[29]. In $SmO_{1-x}F_xFeAs$ a microscopic coexistence of bulk magnetic order and superconductivity and a gradual suppression of $T_N$ was found in the doping range $0.10 < x < 0.17$ [10]. The origin for the difference to our observations in $LaO_{1-x}F_xFeAs$ may be related to a strong electronic coupling to the rare earth magnetic moment in $CeO_{1-x}F_xFeAs$ or to a different level of local structural distortions at the x-dependent crossover from the low temperature orthorhombic lattice structure to the tetragonal phase.

In summary, we present a systematic study of the electronic phase diagram and the respective order parameters of the iron-based superconductor $LaO_{1-x}F_xFeAs$ for x ranging from 0 to 0.20. A first-order-like transition from spin-density wave magnetic order associated with an orthorhombic distortion towards superconductivity is visible in the respective transition temperatures $T_N$, $T_S$ and $T_C$ as well as in the order parameters. This transition occurs for x between 0.04 and 0.05. A quantum-critical point as function of the charge carrier doping is not found in $LaO_{1-x}F_xFeAs$. This competition of the electronic ground states in combination with disorder may be responsible for phase coexistence and phase separation phenomena observed in other iron-pnictides.

**Acknowledgements**
The work at the IFW Dresden and at Cologne University has been supported by the DFG through FOR 538 and SFB 608, respectively.

**Additional information**
Supplementary Information accompanies the paper on www.nature.com/naturematerials.
Reprints and permissions information is available online at http://npg.nature.com/reprintsandpermissions.
Correspondence and requests for materials should be addressed to H.-H. K.

**Competing Interests statement**

The Authors declare that they have no competing financial interests.

**Figure 1** Muon spin relaxation and Mössbauer spectra of $LaO_{1-x}F_xFeAs$.

**a,** Typical zero field µSR spectra. Only for $x \leq 0.04$ a spontaneous muon spin precession indicative for long range ordered magnetism is observed. For $x \geq 0.05$ a paramagnetic signal is observed down to the lowest temperatures. For $x=0.05$ a weak electronic relaxation typical for diluted static magnetism is detected below 5 K in <30% of the signal (visible on the long time scale in the inset). For $x \geq 0.075$ the µSR data prove that no static magnetism is present. **b,** Temperature dependence of the magnetic volume fraction for $x=0$ and $0.04$. Both samples show a transition to a 100% magnetic volume fraction. The ~5% non-magnetic signal is attributed to muons stopping in the sample holder. **c,** Typical transverse field µSR spectra measured in an external field of 0.07 T (for clarity shown in a rotating reference frame with frequency 7.8 MHz). The additional Gaussian relaxation due to the formation of the flux line lattice in the superconducting state is clearly observed below $T_C$. Note that for $x=0.20$ a signal fraction of 15 % does not show this additional relaxation indicating the presence of a non-superconducting volume fraction. **d,** Typical zero field Mössbauer spectra. A magnetic hyperfine splitting is present only for $x \leq 0.04$.

For x=0.05 the paramagnetic line width is slightly increased below 10K due to diluted magnetic correlations. The error bars indicate one standard deviation.

**Figure 2 Structural, magnetic and superconducting order parameter in LaO$_{1-x}$F$_x$FeAs. a,** Temperature dependence of the orthorhombic splitting in LaO$_{1-x}$F$_x$FeAs with x=0.0 and 0.04. The orthorhombic splitting $\varepsilon$ = a-b/a+b is the order parameter of the ferroelastic tetragonal to orthorhombic transition. Orthorhombic lattice constants were determined by the analysis of powder x-ray diffraction patterns through Rietveld refinements. For x=0.04 and x=0.05 the width of the (110)$_{tet}$ peak profile fitted with a single Gaussian is shown for comparison. The single peak transforms into the (020)$_{orth}$/(200)$_{orth}$ pair in the orthorhombic phase. For x=0.04 the structural transition is clearly visible as a step-like increase in the width, whereas the weak temperature dependence of the width observed for x=0.05 unambiguously shows that this sample stays in the tetragonal phase. **b,** Temperature dependence of local magnetic field B$_{muon}$ detecteded by μSR and the Mössbauer magnetic hyperfine field B$_{Hyp}$ for x=0, 0.01, 0.02 and 0.04. Both observables, which are proportional to the ordered magnetic moment size, reveal second order phase transitions and a weak but systematic reduction of the magnetic ordering temperature T$_N$ and of the low temperature saturation value with increasing x. **c,** Temperature dependence of 1/$\lambda_{ab}^2$ which is proportional to the superfluid density n$_s$/m* for all superconducting samples with x ≥ 0.05. The lines represent a power law fit to extract the low temperature value of 1/$\lambda_{ab}^2$. The upturn of the 1/$\lambda_{ab}^2$ below 5K for the strongly underdoped samples does not reflect an increasing superfluid density but is rather a signature of the spurious magnetism present in LaO$_{1-}$

$_x$F$_x$FeAs in this doping range[12]. Note the convex curvature appearing close to T$_C$ for x > 0.10. This might be associated with a distribution of T$_C$ in the sample due to a slightly inhomogeneous F-doping or with multiband effects. The error bars indicate one standard deviation.

**Figure 3 Electronic phase diagram of LaO$_{1-x}$F$_x$FeAs. a,** The doping dependence of the magnetic and superconducting transition temperatures determined from the µSR experiments. Also shown are the tetragonal to orthorhombic structural transition temperatures T$_S$ determined directly from XRD and from susceptibility measurements which show a kink and subsequent strong reduction below T$_S$[8]. **b,** The doping dependence of the low temperature saturation value of the magnetic order parameter B$_{muon}$ (T→ 0) and of the superfluid density n$_s$/m* measured via 1/$\lambda_{ab}^2$ (T→ 0) in TF µSR experiments. The grey data points at x=0.03 and x=0.08 are taken from Carlo et al.[17]. The error bars indicate one standard deviation.

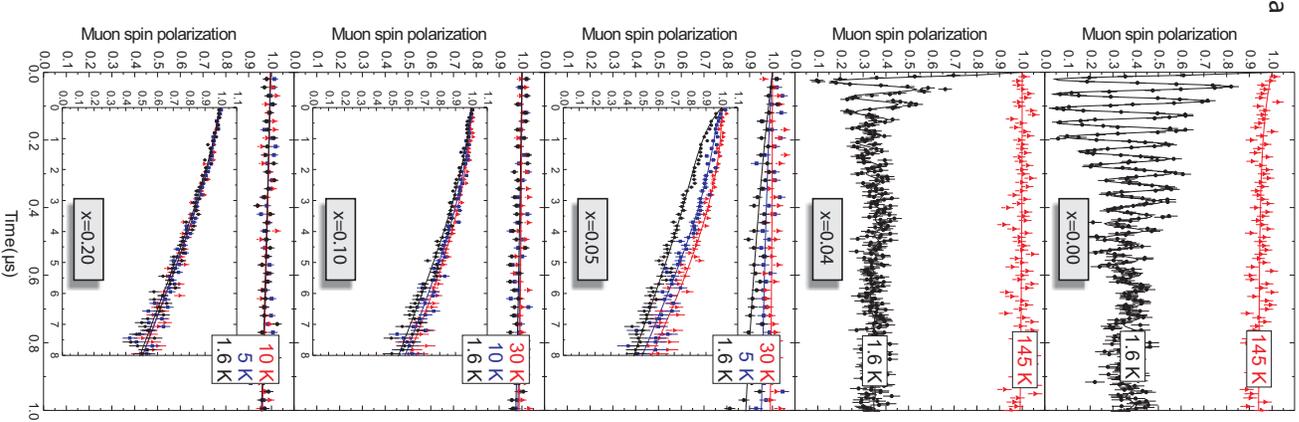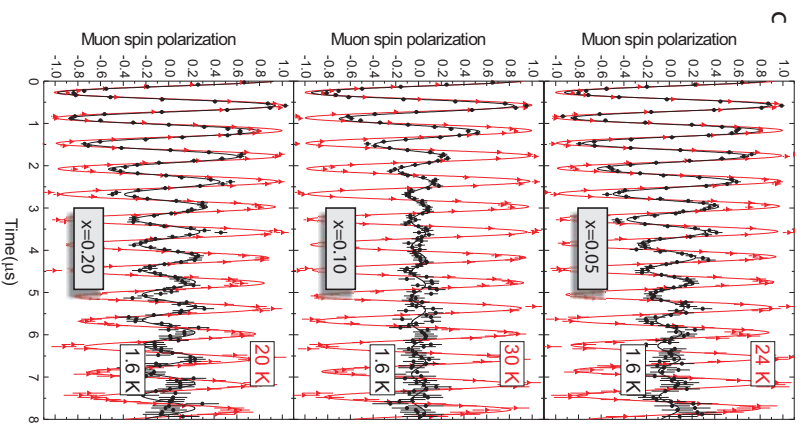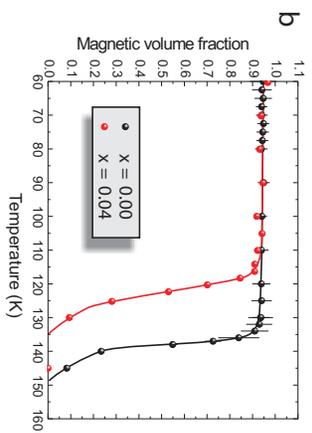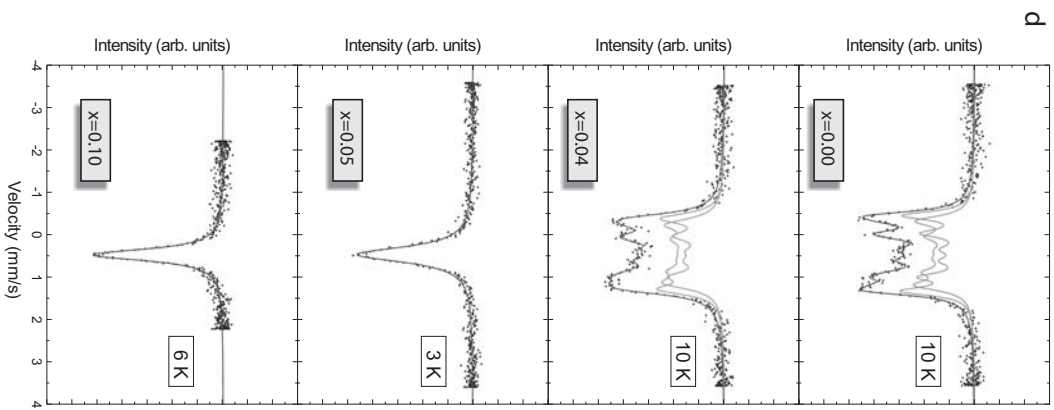

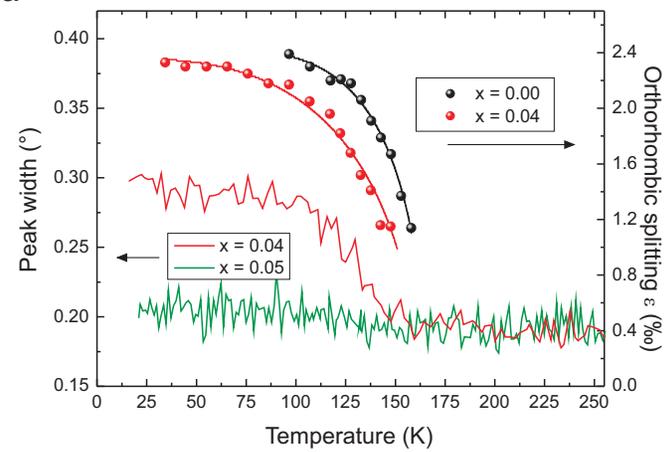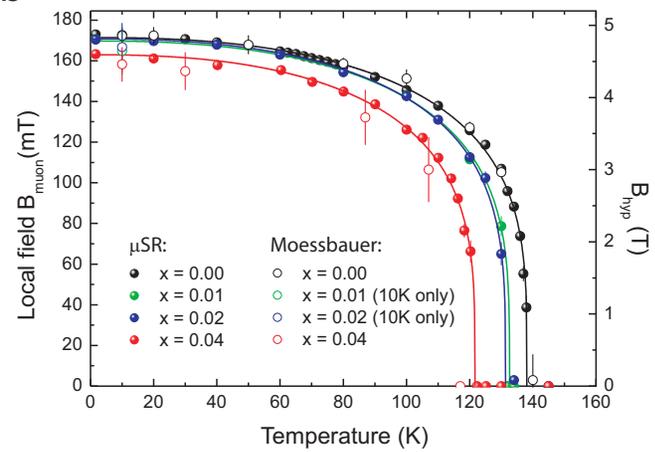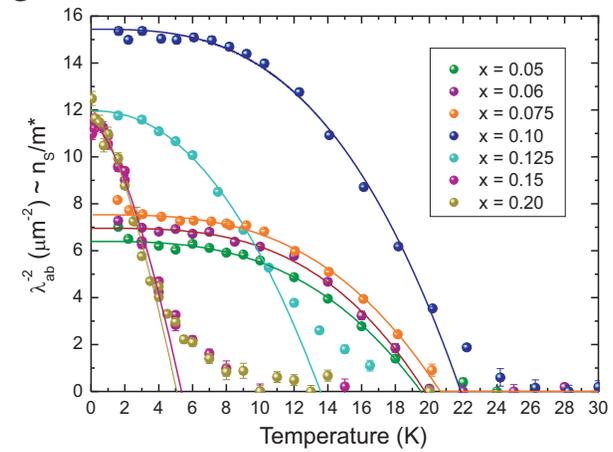

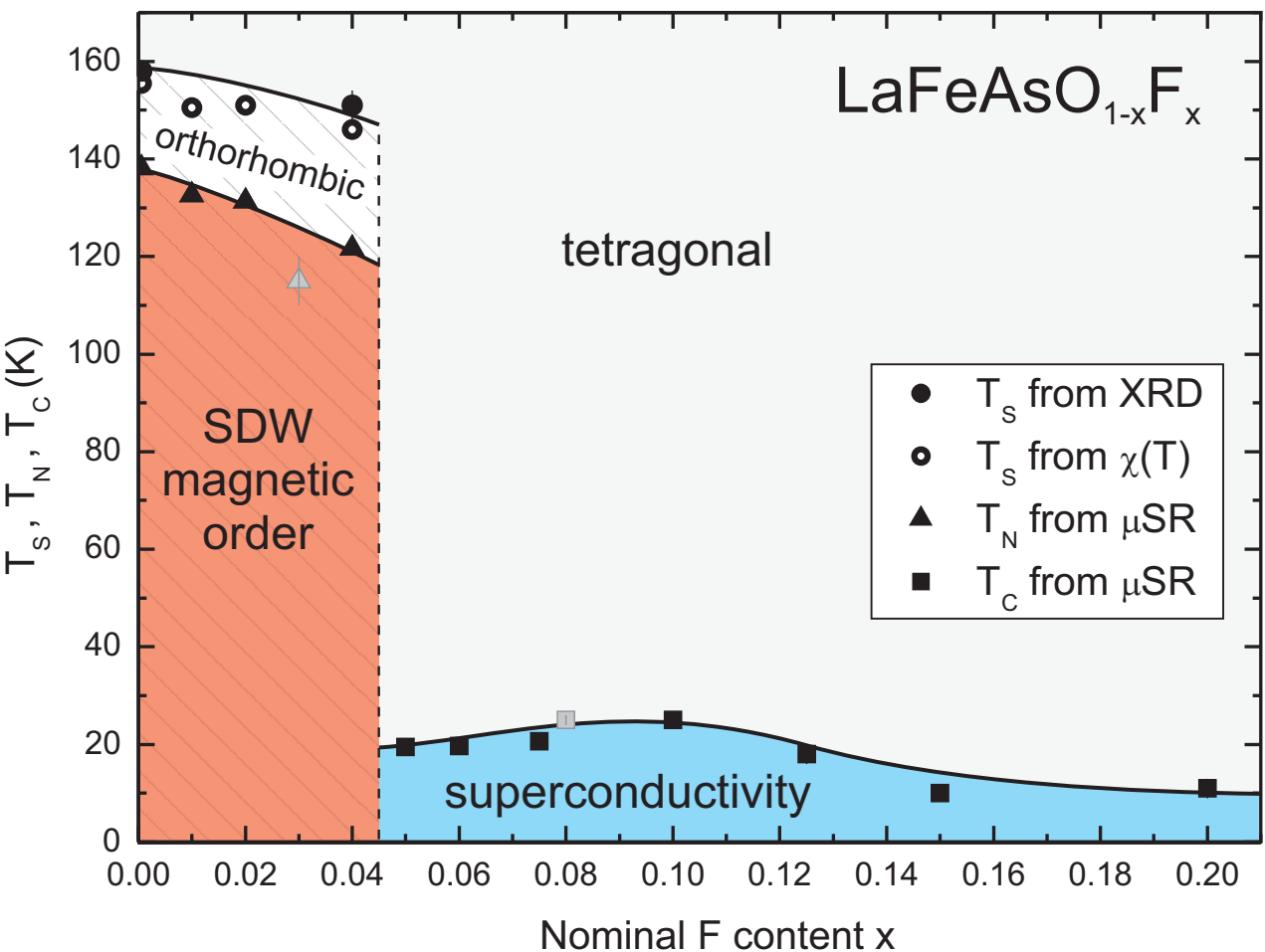

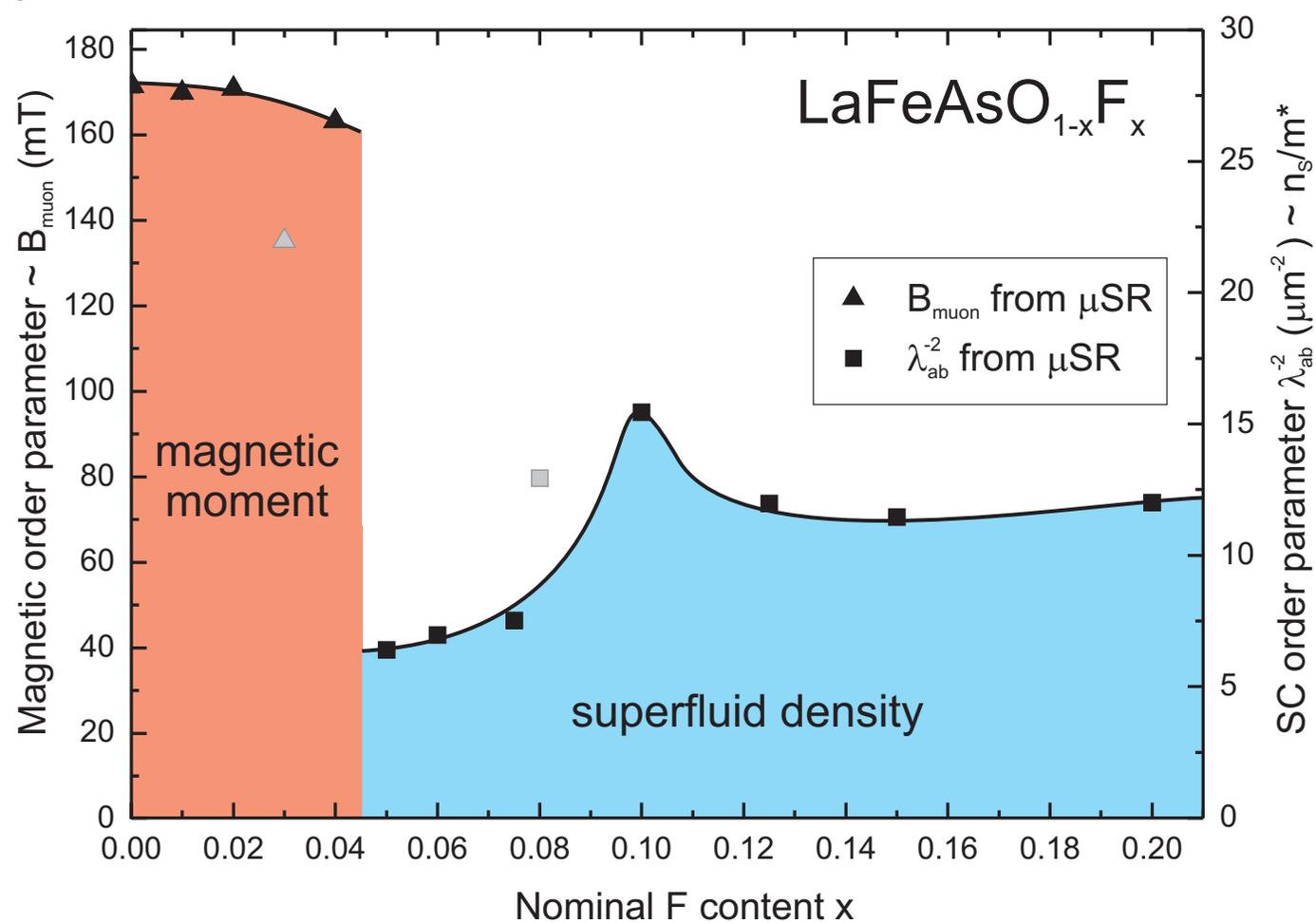

**Supplementary Information for The electronic phase diagram of the LaO$_{1-x}$F$_x$FeAs superconductor**

H. Luetkens, H.-H. Klauss, M. Kraken, F. J. Litterst, T. Dellmann, R. Klingeler, C. Hess, R. Khasanov, A. Amato, C. Baines, M. Kosmala, O.J. Schumann, M. Braden, J. Hamann-Borrero, N. Leps, A. Kondrat, G. Behr, J. Werner and B. Büchner

**1. Sample preparation and characterization**

The samples have been prepared by using a two-step solid state reaction method, similar to that described by Zhu et al.[18], and annealed in vacuum. The crystal structure has been investigated by powder X-ray diffraction. From the X-ray diffraction data impurity concentrations smaller than 1 % are inferred. In addition, we find a continuous decrease of the lattice parameters and the unit cell volume with increasing x consistent with earlier reports[6,9]. Therefore we conclude a homogeneous F doping in the volume of the samples.

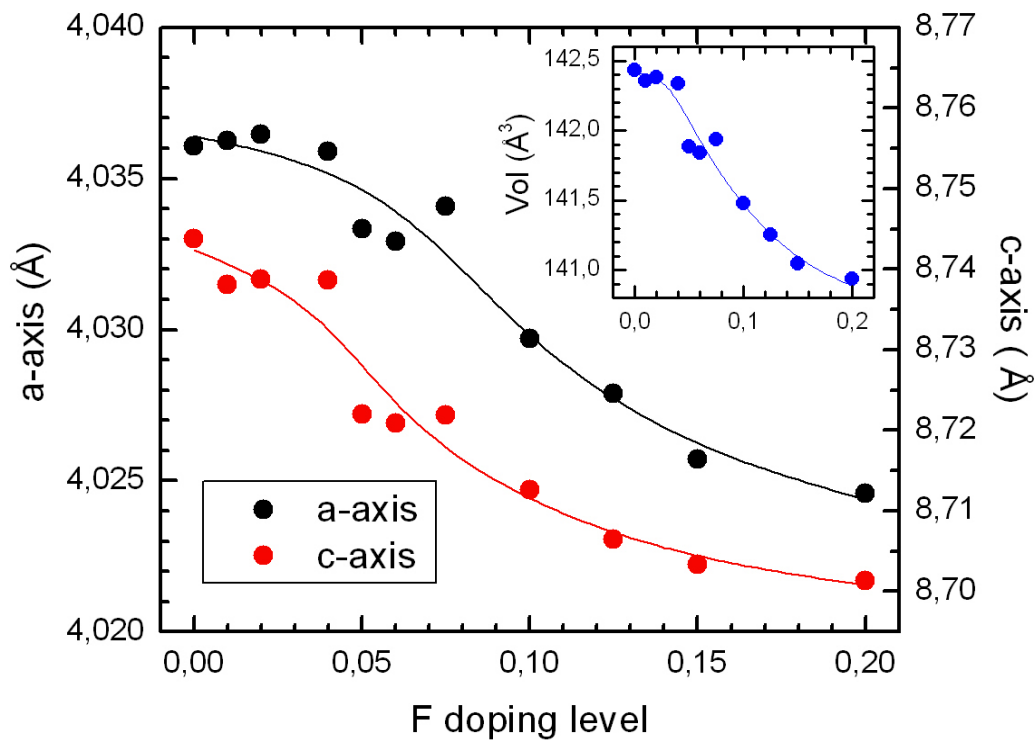

**Figure Sup1. Doping dependence of the lattice constants and cell volume in LaO$_{1-x}$F$_x$FeAs at 300K.**

## 2. Powder x-ray diffraction

LaOFeAs exhibits a structural phase transition at T$_S$ from the tetragonal high-temperature phase (space group P4/nmm) into an orthorhombic low-temperature modification (space group Cmma), with $a_{orth.} \approx b_{orth.} \approx \sqrt{2} \cdot a_{tet.}$ The x-ray powder diffraction profiles at the (110)$_{tet}$ peak position is shown in Fig Sup2a for a high (above T$_S$) and a low temperature (below T$_S$ or base temperature) and x = 0.0, 0.04, 0.05, and 0.10. The (110)$_{tet}$ peak transforms into the (020)$_{orth}$/(200)$_{orth}$ pair in the orthorhombic phase yielding considerable broadening of the peak. While the x=0.04 compound clearly exhibits an increase of the peak width at the tetragonal to orthorhombic transition it is fully absent for the x≥0.05 compounds indicating that for x≥0.05 LaO$_{1-x}$F$_x$FeAs stays in the tetragonal phase down to lowest temperatures.

**Figure Sup2. Doping and temperature dependence of powder x-ray diffraction peaks of LaO$_{1-x}$F$_x$FeAs.** X-ray powder diffraction profiles at the (110)$_{tet}$ peak position for a high (above T$_S$) and a low temperature (below T$_S$ or base temperature) for x = 0.0, 0.04, 0.05, and 0.10. For x=0.0 and x=0.04 the (110)$_{tet}$ peak transforms into the (020)$_{orth}$/(200)$_{orth}$ pair in the orthorhombic phase yielding considerable broadening of the peak. The tetragonal to orthorhombic transition is absent for the x≥0.05 compounds.

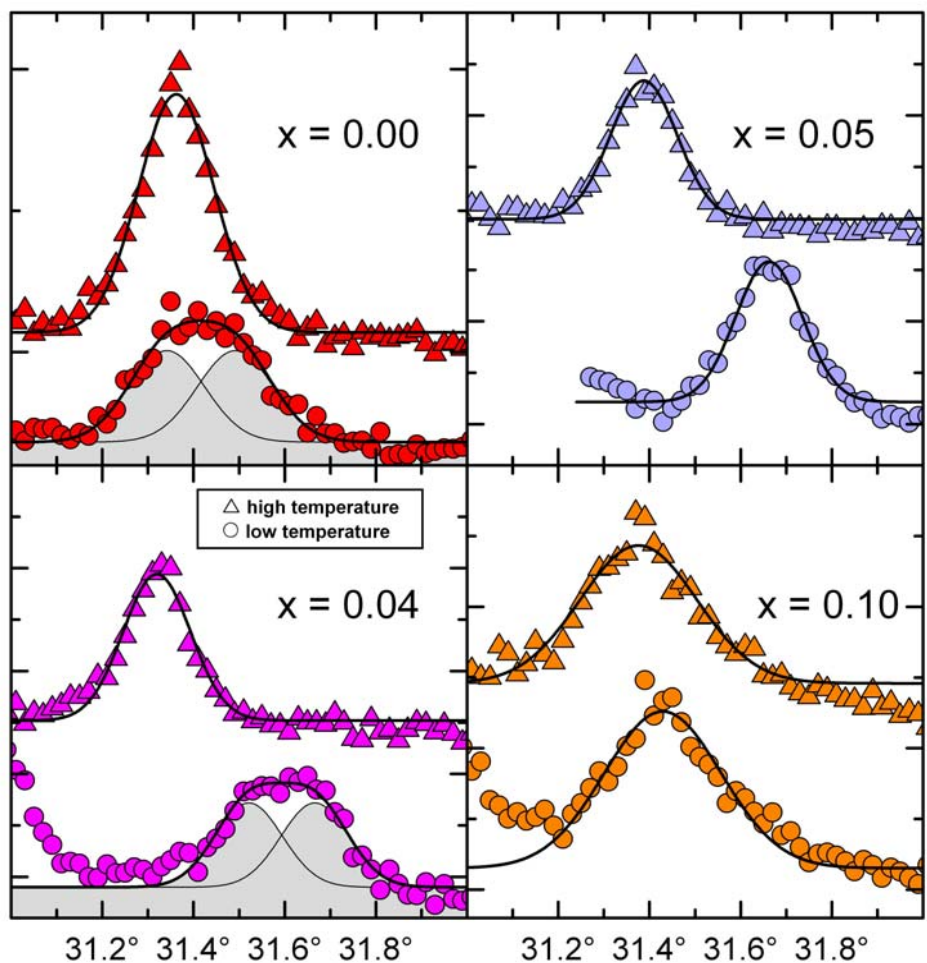

## 2. Muon spin relaxation

Zero field and transverse field (TF) µSR experiments have been performed at the Paul-Scherrer-Institute, Switzerland, using a standard He-flow cryostat and the GPS spectrometer. In Fig. Sup3 a) and b) we present an overview of zero field µSR spectra for different F doping x. In the magnetic regime (x≤0.04) the proper description of the spectra requires a two component relaxation function indicating magnetically different muon sites. Details are reported in[1]. The superconducting samples (x≥0.05) show a typical paramagnetic signal with a tiny Gaussian relaxation due to nuclear moments down to the lowest measured temperature of 1.6 K. No coherent muon spin precession is observed. The underdoped samples (0.05 ≤ x ≤ 0.075) show an additional relaxation below 5 K due to small static internal magnetic fields of electronic origin for a minor fraction of the µSR signal $P_{mag}$ which is 30% for the x=0.05 specimen at 1.6 K and decreases to zero for x=0.10. Since the internal magnetic fields as measured by the precession frequency $f_{muon}$ and the relaxation rate $\lambda_{mag}$ are a factor of 20 smaller in the x=0.05 sample compared to the x=0.04 sample this observation cannot originate from muons within a magnetic cluster exhibiting coherent magnetism. It reflects a depolarization of muons in stray fields in the surrounding of diluted magnetic clusters. Therefore the actual volume fraction of magnetic clusters in our specimens is estimated to be well below 5 % for all temperatures and doping levels x≥0.05. To illustrate the relevance of magnetic inclusions in our superconducting samples we plot the average local magnetic field density $P_{mag}*f_{muon}$ and $P_{mag}*\lambda_{mag}$ in Fig. Sup3 c) for 0≤x≤0.2. There is no relevant magnetic field density in the superconducting samples of $LaO_{1-x}F_xFeAs$. This is very different from recent studies on $SmO_{1-x}F_xFeAs$ [2] and observations in high-$T_C$-cuprates where a coexistence of magnetism and superconductivity has been observed in the full sample volume of underdoped $La_{2-x}Sr_xCuO_4$ and $Y_{1-x}Ca_xBa_2Cu_3O_6$ [3].

**Figure Sup3. Temperature dependent ZF-µSR-spectra of LaO$_{1-x}$F$_x$FeAs.**
**a,** For x= 0, 0.01, 0.02 and 0.04 a spontaneous muon spin precession is found below the respective Néel temperatures. **b,** For x=0.05, 0.075, 0.10, 0.125, 0.15 and 0.20 no spontaneous precession is observed. Only for x=0.05 and 0.075 a small signal fraction (<30 %) exhibits a weak relaxation of electronic origin below 5 K (note the 6 times longer time scale than in **a**. We attribute this relaxation to diluted magnetic clusters within the major superconducting volume due to local variations of the F content. **c,** Local magnetic field density $P_{mag}*f_{muon}$ and $P_{mag}*\lambda_{mag}$ as a function of F content at T=1.6 K proving that no bulk coexistence of magnetism and superconductivity is observed in LaO$_{1-x}$F$_x$FeAs. The error bars indicate one standard deviation.

## 3. Mössbauer spectroscopy

The Mössbauer spectroscopy experiments have been done using a standard He-flow cryostat [source: $^{57}$Co-in-Rh matrix at room temperature; emission line half width at half maximum: 0.130(2) mm/s].

Temperature dependent Mössbauer spectra of LaO$_{1-x}$F$_x$FeAs with x= 0.04 and 0.05 are shown in Fig. Sup4. For x=0.04, above 140 K the spectra can be fitted with a single Lorentzian line with an isomer shift of S=0.52(1) mm/s. This is in the typical range of low or intermediate spin Fe(II). Below 140 K, a splitting of the absorption line reveals the formation of a magnetic hyperfine field. In this temperature range the spectra have been analyzed by diagonalizing the hyperfine Hamiltonian including electric quadrupole and magnetic hyperfine interaction. In order to properly describe the spectra we have used two sextets of equal intensities which technically account for the broadening. Both sextets show an electric quadrupole splitting of QS~0.3 mm/s due to a small deformation of the FeAs$_4$ tetrahedron below the structural phase transition[1]. The main result is the observation of a magnetic hyperfine field with a mean saturation value at low temperatures of B$_{hyp}$(0)= 4.46(24) T which is ~9% smaller than in undoped LaOFeAs[1].

In contrast, the spectra for x=0.05 show no magnetic splitting. An analysis using a single Lorentzian line reveals a gradual increase of the linewidth from its high-temperature value of 0.17 mm/s to 0.24 mm/s below 10 K pointing to very dilute magnetic correlations in this sample in agreement with the μSR results discussed above.

**Figure Sup4. Temperature dependent Mössbauer spectra of $LaO_{1-x}F_xFeAs$ with x= 0.04 and 0.05.**

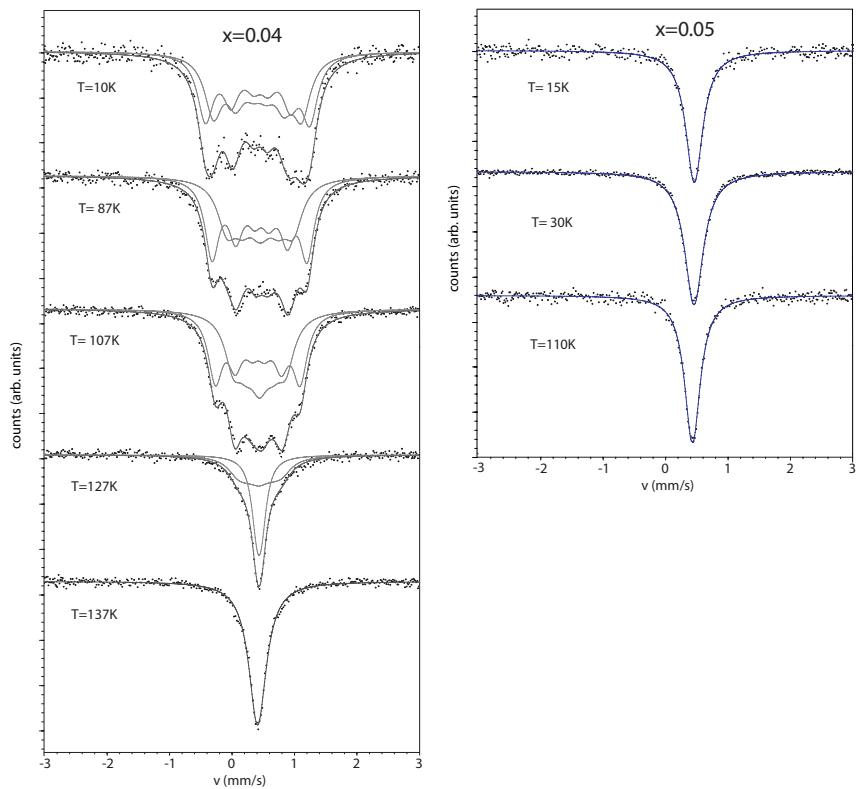